\documentclass[preprint,12pt]{revtex4}
\usepackage{graphicx}
\def\beq{\begin{equation}}
\def\eeq{\end{equation}}
\def\bea{\begin{eqnarray}}
\def\eea{\end{eqnarray}}

\begin{document}
\title{ Higgs Boson Searches via Dileptonic Bottomonium Decays }
\author{Saime Solmaz}
\email{skerman@balikesir.edu.tr} \affiliation{Bal{\i}kesir
University, Physics Department, TR10100,
Bal{\i}kesir, Turkey}%
\date{\today}
\begin{abstract}
We explore  the pseudoscalar $\eta_b$ and the scalar $\chi_{b0}$
decays into $\ell^+\ell^-$ to probe whether it is possible to probe
the Higgs sectors beyond that of the Standard Model. We, in
particular, focus on the Minimal Supersymmetric Standard Model, and
determine the effects of its Higgs bosons on the aforementioned
bottomonium decays into lepton pairs. We find that the dileptonic
branchings of the bottomonia can be sizeable for a relatively light
Higgs sector.

\end{abstract}
\maketitle

\section{Introduction}

Having the LHC started, the search for physics at the terascale
has entered a new phase. The ATLAS and CMS experiments at the
LHC will search for new particles and forces while LHCb will
provide a more accurate description of flavor physics. Each
experiment, combined with others, will provide important
information about nature of new physics awaiting discovery. It
is thus rather timely to discuss and analyze ways of extracting
${\rm TeV}$ scale physics in relation to the measurements at
the LHC experiments.

In terms of its content and goal, the present work falls in the
interface between flavor physics and Higgs physics in that we
aim at exploring finger prints of yet-to-be discovered Higgs
sector (to be discovered at the CMS and ATLAS experiments
\cite{cmsatlas}) in the leptonic decay distributions of heavy
hadrons (to be accurately measured at the LHCb experiment
\cite{lhcb}).

At present, we do not have any clue of what Higgs sector is
awaiting for discovery at the LHC. On the other hand, the
experiments at $B$ factories have, by now, established a grand
view of the flavor physics. The experimental precision is
increasing steadily and has already started challenging our
understanding of the flavor violation. Over the years, various
$B$ meson decay rates and charge asymmetries have been measured
and novel quarkonium states have been discovered. The $B$ meson
inventory of the existing storage rings comes from the decays
of $(b\,\overline{b})$ states (bottomonium states) produced at
asymmetric electron-positron collisions ($e.g.$ PEPII at SLAC
and    KEK-B at KEK). Of course, all kinds of bottomonia with
varying spin and CP quantum numbers will be produced at the LHC
in gluon-gluon or gluon-gluon-gluon fusion channels.

In principle, one ought to use every single opportunity to
extract information about other sectors of a given theory by
using the available information from $B$ physics. Examples of
such efforts involve quark EDMs \cite{edm} and flavor-violation
Higgs connection \cite{flavor-higgs}. The radiative, leptonic
or semileptonic decays of hadrons are particularly suitable for
strengthening experimental identification and theoretical
prediction, and thus,  in this work we attempt at answering the
following question: {\bf By measuring the decay rates of
certain    $(b\, \overline{b})$ states, preferably but not
necessarily    into $\ell^+\ell^-$, can we establish the
existence and nature of Higgs bosons?} The choice of
bottomonium system stems from not only its perturbative nature
but also its appreciable coupling to Higgs fields.

In what follows, in regard to the question raised above, we will
study a generic Higgs sector extending that of the SM. In the next
section, we will provide an explicit discussion of the dileptonic
Bottomonium decays into lepton pairs. In Sec. 3 we will numerically
analyze the decay rates by taking MSSM to be the new physics
candidate model and fixing the unknown parameters to two different
data  sets taken from LEP indications and from SPS1a point. In Sec.4
we conclude.

\section{Dileptonic Bottomonium Decays}

The quarkonium systems have been under intense study since
    the discovery of the charm quark \cite{charm}. That light
    {\rm MeV}--mass Higgs bosons could be produced in
    quarkonium decays was first discussed in
    \cite{ilkler,hunthiggs}. The decays of additional {\rm
    TeV}--mass heavy quark bound states into fermions as well
    as Higgs and gauge bosons have been analyzed in
    \cite{barger1}. In this work we will discuss Higgs boson search via
    dileptonic bottomonium decays. The two sides, hadronic and Higgs
    aspects,  of our discussions can be described as follows:
\begin{enumerate}
\item We will focus on bottomonium states, in particular, the
pseudoscalar $\eta_b$  (an $S$-wave $J^{P C} = 0^{-+}$ state) and
the scalar $\chi_{b0}$ (a $P$-wave $J^{PC}= 0^{++}$ state). Unlike
the charmonium system where such states have already been
experimentally established, the experimental efforts still
continue to establish quantum numbers of $\eta_b$ and $\chi_{b 0}$
though they have already been observed
\cite{etabexp,chibexp,etachitheor,pdg}.  The experiments at
Tevatron, $B$ factories and LHC are expected to fully construct
and measure partial widths of these $b\, \overline{b}$ mesons.

\item We will not restrict ourselves to standard model
    Higgs sector. In fact, we will consider models with two
    Higgs doublets, $H_u$ and $H_d$ one giving mass to
    down-type fermions other to up-type fermions, as
    encountered in the MSSM. (One may, of course, consider
    more general Yukawa structures \cite{atwood}.) The
    spectrum consists of three Higgs bosons: the CP-even
    ones $h$ and $H$ and a CP-odd one $A$. Their
    interactions with $b$ quark  and charged leptons are
    given by
\begin{eqnarray}
-{\cal{L}}_{higgs} &=& g_{h}^f\, \overline{f}\, f\, h + g_{H}^f\,
\overline{f}\, f\, H + g_{A}^f\, \overline{f}\,i\gamma_5\, f\, A
\end{eqnarray}
where $f=b,\ell$, and the Yukawa couplings $g^{f}_X$ are
given by
\begin{eqnarray}
g_{h}^{f} &=& h_f^{SM}\, \left[\sin(\beta-\alpha)-\tan\beta\,
\cos(\beta-\alpha)\right] \nonumber\\
g_{H}^{f} &=& h_f^{SM}\, \left[\cos(\beta-\alpha)+\tan\beta\,
\sin(\beta-\alpha)\right]\nonumber\\
g_{A}^{f} &=& h_f^{SM}\, \tan\beta\,.
\end{eqnarray}
Here $\tan\beta = \langle H_u^0\rangle / \langle H_d^0\rangle$,
$\alpha$ is the mixing between $H_u^0-\langle H_u^0\rangle$ and
$H_d^0 - \langle H_d^0 \rangle$ such that
$\alpha=1/2\arcsin\left[-(m^2_A+m^2_Z)/(m^2_H-m^2_h)\sin2\beta\right]$,
$h_f^{SM} = (g_2 m_f)/(2 M_W)$ is the Yukawa coupling of fermion $f$
in the SM. If there exists explicit CP violation sources in the
theory then none of the Higgs bosons can possess definite CP quantum
number, and thus they couple to fermions as $\overline{f}(a + i b
\gamma_5) f$ as in, for instance, the MSSM with complex soft terms
with one-loop Higgs potential \cite{demir}.
\end{enumerate}

Having specified the framework in both Higgs and meson sides,
we now turn to an explicit computation of the decay rates of
bottomonia. In this respect, the decay rates of $\eta_b$ and
$\chi_{b 0}$ into lepton pairs are then given by
\begin{eqnarray}
\label{formula}
\Gamma\left(\eta_b \rightarrow \ell^+ \ell^-\right) &=& \frac{3}{8
\pi^2}\, \frac{\left|R_S(0)\right|^2}{M_{\eta_b}^2}\, \beta_{\ell} \nonumber\\
&\times& \left\{\left(\frac{g_A^b g_A^{\ell}}{1-r_A}\right)^2 + 4
r_{\ell} \left(\frac{g_Z^2}{1-r_Z}\right)^2 + 4 \sqrt{r_{\ell}}
\frac{g_Z^2 g_A^b g_A^\ell}{(1-r_A) (1-r_Z)}\right\}\nonumber\\
\Gamma\left(\chi_{b 0}\rightarrow \ell^+ \ell^-\right)  &=&
\frac{27}{8 \pi^2}\, \frac{\left|R_P^{\prime}(0)\right|^2}{M_{\chi_{b 0}}^4}\, \beta_{\ell}^3 \nonumber\\
&\times& \left(\frac{g_h^b g_h^\ell}{1-r_h}+ \frac{g_H^b
g_H^\ell}{1-r_H}\right)^2
\end{eqnarray}
where $R_S(0)$ ($R_P^{\prime}(0)$) is the S-wave (derivative of
P-wave) quarkonium wavefunction at the origin \cite{barger1},
$g_Z= e/(4 \sin\theta_W \cos\theta_W)$, $r_i = m_i^2/M_X^2$
($i=\ell, A,h,H,Z$), and $\beta_{\ell} = \left(1- 4
m_{\ell}^2/M_{X}^2\right)^{1/2}$ where $X=\eta_b$ for $\eta_b
\rightarrow \ell^+ \ell^-$ and $X=\chi_{b 0}$ for $\chi_{b 0}
\rightarrow \ell^+ \ell^-$.

From these decay rates one notes that:
\begin{enumerate}
\item Thanks to their $J^{PC}$ structures, the two
    bottomonia, $\eta_{b}$ and $\chi_{b 0}$, explicitly
    distinguish between CP=+1 and CP=-1 Higgs bosons. This
    aspect proves very important for establishing the
    nature of the Higgs bosons as well as  structure of the
    non-SM Higgs sector at the LHC and its successor NLC
    (see \cite{edm} for a detailed discussion of different
    $J^{PC}$ mesons).

\item The couplings of the Higgs bosons to down-type
    fermions experience big enhancements at large
    $\tan\beta$ as preferred by LEP experiments. Indeed,
    contributions of $A$ and $H$ grow as
    $\left(\tan\beta/M_{H,A}\right)^2$ which can provide a
    {\it detectable} signal for collider experiments such
    as the LHCb.

\item As is seen from (\ref{formula}), as a direct
    consequence of the quantum numbers of $\eta_b$ meson,
    the decay $\eta_{b}\rightarrow \ell^+ \ell^-$
    exclusively involves the vector bosons and pseudoscalar
    Higgs bosons. On the other hand, again due to its
    quantum numbers, the decay $\chi_{b 0}\rightarrow
    \ell^+ \ell^-$ singles out the CP=+1 Higgs bosons.
    (Nevertheless, one keeps in mind that  $\chi_{b
    0}\rightarrow \ell^+ \ell^-$ can exhibit a
    non-negligible dependence on the masses of the
    pseudoscalar Higgs bosons depending on the details of
    the mixing angle $\alpha$ of the CP-even Higgs sector).
    This prime difference between the two mesons proves
    highly useful for probing the nature of the `new
    physics' that will be discovered at the LHC. Depending
    on the nature of the deviations from the SM
    expectations, one might determine, within the
    experimental uncertainities, whether the new physics
    involve new pseudoscalar Higgs bosons (like the MSSM or
    NMSSM) or new scalar Higgs bosons (like MSSM or
    U(1)$^{\prime}$ models or NMSSM) or new gauge bosons
    (like U(1)$^{\prime}$ invariance of left-right
    symmetric models).

\item In (\ref{formula}) we have focused particularly on
    extended Higgs sectors (taken to be a generic
    two-doublet model fitting to the Higgs sector of the
    MSSM). However, one can consider extended gauge sectors
    as well. In this case, similar to the $Z/W$ boson
    contributions, one expects anomalous behavior in
    $\eta_b \rightarrow \ell^+ \ell^-$ (compared to the SM
    prediction) to arise also from extended gauge sectors
    containing $Z^{\prime}/W^{\prime}$ gauge bosons. In
    this work we will not investigate this option sice
    experimental bounds force $Z^{\prime}/W^{\prime}$ to
    stay heavy (though in realistic models the Higgs sector
    itself behaves differently \cite{levent}).

\item In the decoupling limit \cite{hunthiggs,demir}, it
    turns out that $\beta-\alpha \sim \pi/2$ in which case
    $h$ behaves as in the SM yet $H$ and $A$ Higgs bosons
    possess $\tan\beta$--enhanced Yukawa interactions.

\end{enumerate}
In the next section we will perform a numerical study of the
decay rates (\ref{formula}) in view of disentangling $H$ and
$A$ effects from the rest. The analysis, once confirmed
experimentally, might provide important information about the
nature of the Higgs sector awaiting discovery at the LHC.

\section{Numerical Analysis}

In this section we analyze the decay widths discussed above
numerically. In doing this, the SM prediction for the decay
rate will be compared with those of the MSSM for $\chi_{b0}$
and $\eta_{b}$ decays, comparatively. In particular, we take
Higgs boson of the SM degenerate in mass with the lightest
Higgs boson of the MSSM, and  consider the ratios
\begin{eqnarray}
\label{formula2} \frac{\Gamma^{\rm{MSSM}}\left(\eta_b \rightarrow
\ell^+ \ell^-\right)}{\Gamma^{\rm{SM}}\left(\eta_b \rightarrow
\ell^+ \ell^-\right)} = 1+\left(\frac{g_A^b\,
g_A^{\ell}}{g^2_Z}\right)^2 \frac{1}{4 r_{\ell}}
\left(\frac{1-r_Z}{1-r_A}\right)^2 +\left(\frac{g_A^b\,
g_A^{\ell}}{g^2_Z}\right) \frac{1}{\sqrt{r_{\ell}}}
\left(\frac{1-r_Z}{1-r_A}\right)
\end{eqnarray}
and
\begin{equation}
\label{formula3} \frac{\Gamma^{\rm{MSSM}}\left(\chi_{b 0}\rightarrow
\ell^+ \ell^-\right)}{\Gamma^{\rm{SM}}\left(\chi_{b 0}\rightarrow
\ell^+ \ell^-\right)} =\left(\frac{g_h^b\,g_h^{\ell}}{h_b^{SM} \,
h_{\ell}^{SM}}\right)^2+\left(\frac{g_H^b\,g_H^{\ell}}{h_b^{SM} \,
h_{\ell}^{SM}}\right)^2 \left(\frac{1-r_h}{1-r_H}\right)^2 +2 \frac
{\left(g_h^b\, g_h^{\ell}\, g_H^b\,
g_H^{\ell}\right)}{\left(h_b^{SM} \, h_{\ell}^{SM}\right)^2}
\left(\frac{1-r_h}{1-r_H}\right)\nonumber\\
\end{equation}
in making the numerical estimates. The parameter values for
which these ratios exceed unity significantly are expected to
yield observable signals. In course of the analysis we scan the
parameter space of the MSSM Higgs sector by varying  $m_h$,
$m_H$, $m_A$ and $\tan\beta$ in a considerably wide range. We
focus on two parameter ranges:
\begin{itemize}
\item {\texttt{SUSY Parameter Space I (SPSI)}}:
\begin{eqnarray}
m_h=98\pm5\rm{\,GeV}, m_H=115\pm5\rm{\,GeV}, m_A=89\pm5\rm{\,GeV}, \tan\beta=10\pm2.5
\end{eqnarray}
which is inspired from the reanalysis of the LEP results mentioned
in \cite{levent}.

\item {\texttt{SUSY Parameter Space II (SPSII)}}:
\begin{eqnarray}
m_h=115\pm5\rm{\,GeV}, m_H=425\pm5\rm{\,GeV},
m_A=424.9\pm5\rm{\,GeV}, \tan\beta=10\pm2.5
\end{eqnarray}
which is inspired from the SPS1a parameter space of the MSSM.
\end{itemize}

In plotting a particular figure we vary one parameter while keeping
the rest at their mid-values. Depicted in Fig. 1 (for
$\texttt{SPSI}$) and Fig. 2 (for $\texttt{SPSII}$) are the ratios in
(\ref{formula2}) and (\ref{formula3}) as a  function of the lightest
Higgs boson mass $m_h$.
\begin{figure}[htb]
\begin{center}
\vspace{0.5cm}
    \includegraphics[height=5.5cm,width=16cm,angle=0]{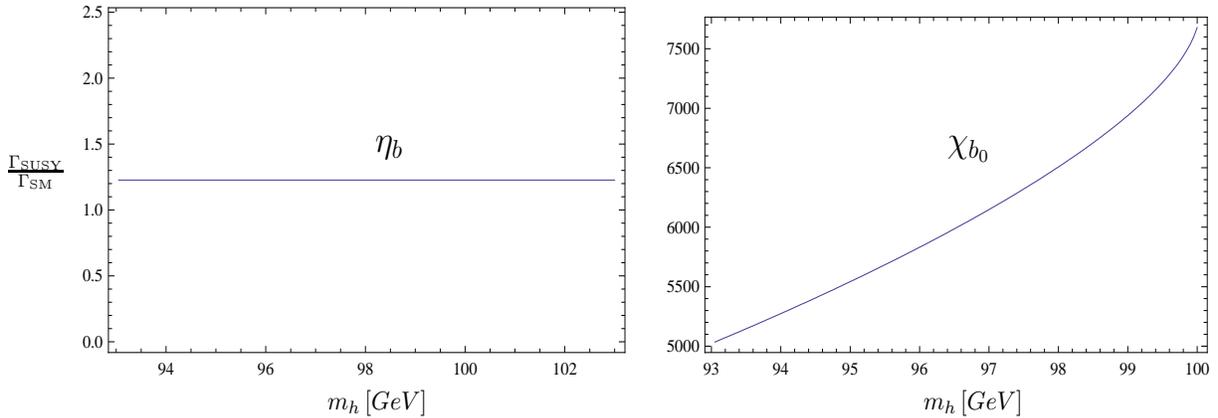}\hspace{1.5 cm}
    \vspace{-.5cm}
    \caption[]{Variation of the decay rate ratios against the lightest CP-even Higgs mass  $m_h$ for
     $\eta_b$ (left panel) and $\chi_{b0}$ (right panel), for the $\texttt{SPSI}$ parameter space.}
    \label{fig1}
    \end{center}
\end{figure}
As can be seen from the left panels of the figures, the ratio of the
SUSY prediction to the SM prediction does not vary for the $\eta_b$
decay and these ratios are approximately $1.23$ and $1.02$ for the
Figs. \ref{fig1} and  \ref{fig2}, respectively. This  can be easily
understood from (\ref{formula2}), to which the CP-even Higgs bosons
do not contribute at all.  The $\eta_b$ decay would probe new CP=-1
Higgs bosons and new gauge bosons, as can be seen from the same
equation. Nevertheless, the difference persistent in the MSSM's
prediction can be an important clue for the future measurements.

While the impact of different  data sets  are sensible for the
$\eta_b$ decay, for the $\chi_{b0}$ decay it turns out to be much
stronger as can be seen from the right panels of the same figures.
For instance the impact of increasing the mass of the lightest
CP-even Higgs boson can enhance the SUSY/SM ratio from $5050$ to
$7700$ for $\texttt{SPSI}$ parameter set. Similarly, it increases
from $62.5$ to $84.5$  for $\texttt{SPSII}$ set. Here, as a result
of the quantum numbers of $\chi_{b 0}$, the contributions of the $h$
and $H$ bosons become clearly visible, which can be seen from
(\ref{formula3}). Normally, as $m_h$ increases the $r_h$, and hence,
the decay rate ratios increase but the dominant behavior is
determined by the coupling terms. In any case, the prediction of the
MSSM is very large than that of the SM prediction, which makes the
$\chi_{b0}$ decay a promising candidate for probing the new CP-even
Higgs bosons of the `new physics'.

\begin{figure}[htb]
\begin{center}
\vspace{0.5cm}
    \includegraphics[height=5.5cm,width=16cm,angle=0]{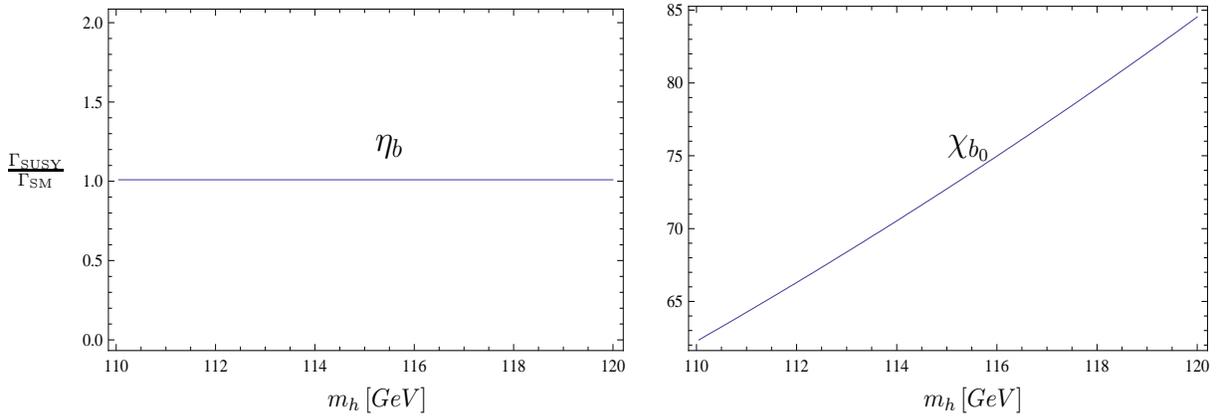}\hspace{1.5 cm}
    \vspace{-.5cm}
    \caption[]{The same as Fig.\ref{fig1}, but for $\texttt{SPSII}$ parameter set.
    }
    \label{fig2}
    \end{center}
\end{figure}

\begin{figure}[htb]
\begin{center}
\vspace{0.5cm}
    \includegraphics[height=5.5cm,width=16cm,angle=0]{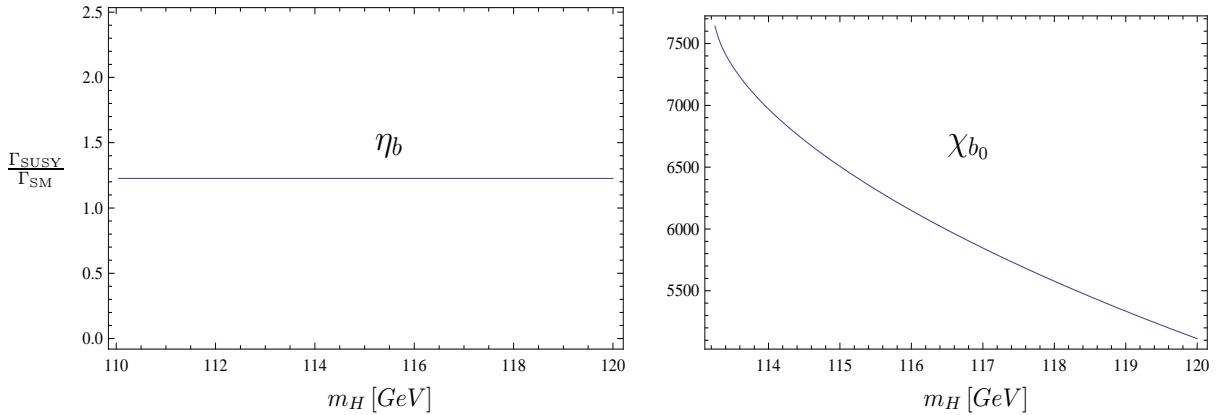}\hspace{1.5 cm}
    \vspace{-.5cm}
    \caption[] {Variation of the decay rate ratios against the heavy CP-even Higgs mass  $m_H$ for
     $\eta_b$ (left panel) and $\chi_{b0}$ (right panel), for the $\texttt{SPSI}$ parameter space.}
    \label{fig3}
    \end{center}
\end{figure}

\begin{figure}[htb]
\begin{center}
\vspace{0.5cm}
    \includegraphics[height=5.5cm,width=16cm,angle=0]{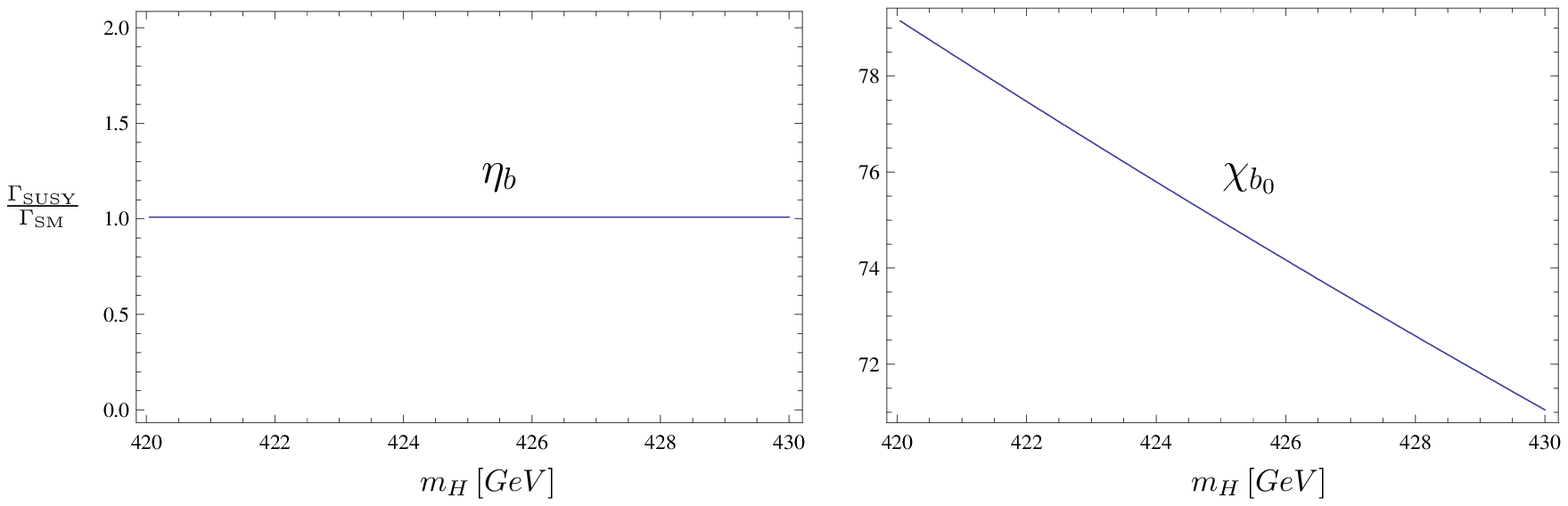}\hspace{1.5 cm}
    \vspace{-.5cm}
    \caption[] {The same as Fig.\ref{fig3}, but for $\texttt{SPSII}$ parameter set.
    }
    \label{fig4}
    \end{center}
\end{figure}

Depicted in  Figs. \ref{fig3} and  \ref{fig4}, are variations of the
decay rate ratios with the heavy CP-even Higgs boson mass, $m_H$.
The general behavior is similar to those in Figs. \ref{fig1} and
\ref{fig2}, except that the $\chi_{b0}$ decay ratio
  decreases as  $m_H$ increases, for
 both of the parameter sets. This can be understand from (\ref{formula3})
 wherein the decay rate ratio is inversely proportional to  ${m_H}^2$.

\begin{figure}[htb]
\begin{center}
\vspace{0.5cm}
    \includegraphics[height=5.5cm,width=16cm,angle=0]{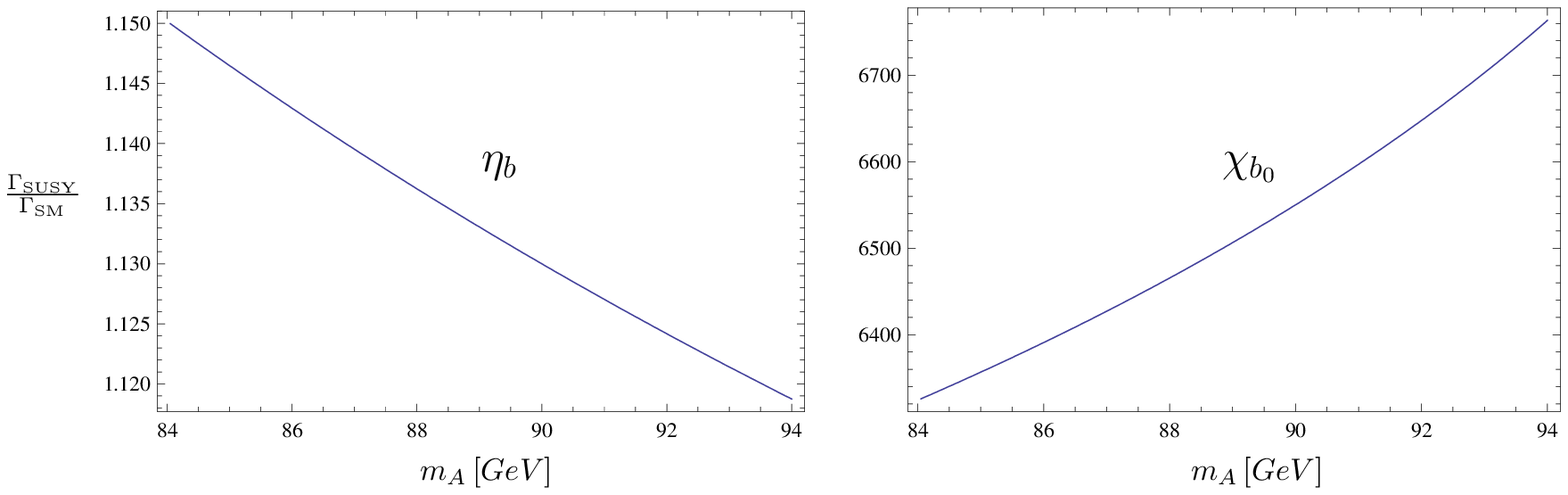}\hspace{1.5 cm}
    \vspace{-.5cm}
    \caption[] {Variation of the decay rate ratios against the CP-odd Higgs boson mass  $m_A$ for
     $\eta_b$ (left panel) and $\chi_{b0}$ (right panel), for the $\texttt{SPSI}$ parameter space.}
    \label{fig5}
    \end{center}
\end{figure}

\begin{figure}[htb]
\begin{center}
\vspace{0.5cm}
    \includegraphics[height=5.5cm,width=16cm,angle=0]{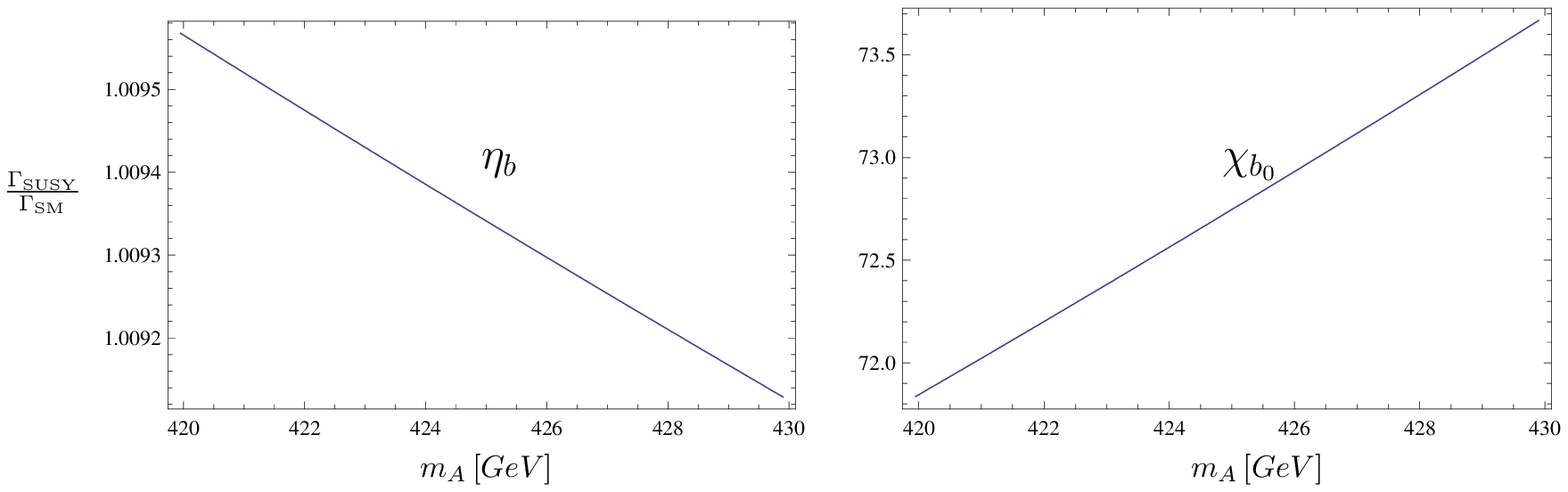}\hspace{1.5 cm}
    \vspace{-.5cm}
    \caption[] {The same as Fig.\ref{fig5}, but for $\texttt{SPSII}$ parameter set.
    }
    \label{fig6}
    \end{center}
\end{figure}

Depicted in Figs. \ref{fig5} and  \ref{fig6} are the ratios of the
decay rates as a function of the pseudoscalar Higgs boson mass
$m_A$. As suggested by the left panels of the figures be (the
$\eta_b$ decays) the ratios decrease with increasing $m_A$ to a
small extend, for both of the parameter sets. For the $\chi_{b 0}$
decays, the reaction response to variation in $m_A$ is much more
pronounced: The decay rate ratio ranges from $6325$ to  $6760 $ for
$\texttt{SPSI}$, and does from $71.85$ to $73.70$ for
$\texttt{SPSII}$. It is important to stress that, as suggested by
formulae (\ref{formula3}), the $m_A$ dependence of the $\chi_{b 0}$
decay follows from the dependencies of the $H$ and $h$ couplings on
the Higgs mixing angle $\alpha$.

\begin{figure}[htb]
\begin{center}
\vspace{0.5cm}
    \includegraphics[height=5.5cm,width=16cm,angle=0]{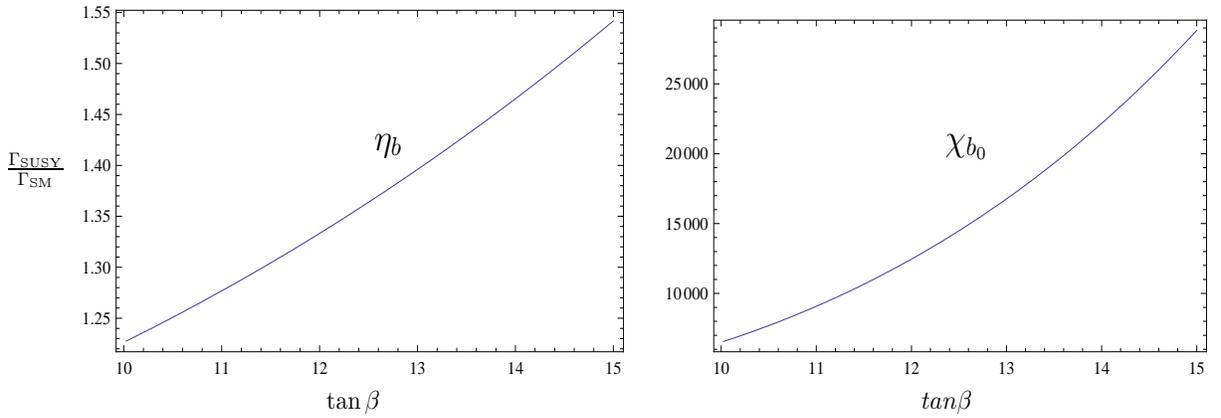}\hspace{1.5 cm}
    \vspace{-.5cm}
    \caption[] {Variation of the decay rate ratios against $\tan\beta$  for
     $\eta_b$ (left panel) and $\chi_{b0}$ (right panel), for the $\texttt{SPSI}$ parameter space.}
    \label{fig7}
    \end{center}
\end{figure}

\begin{figure}[htb]
\begin{center}
\vspace{0.5cm}
    \includegraphics[height=5.5cm,width=16cm,angle=0]{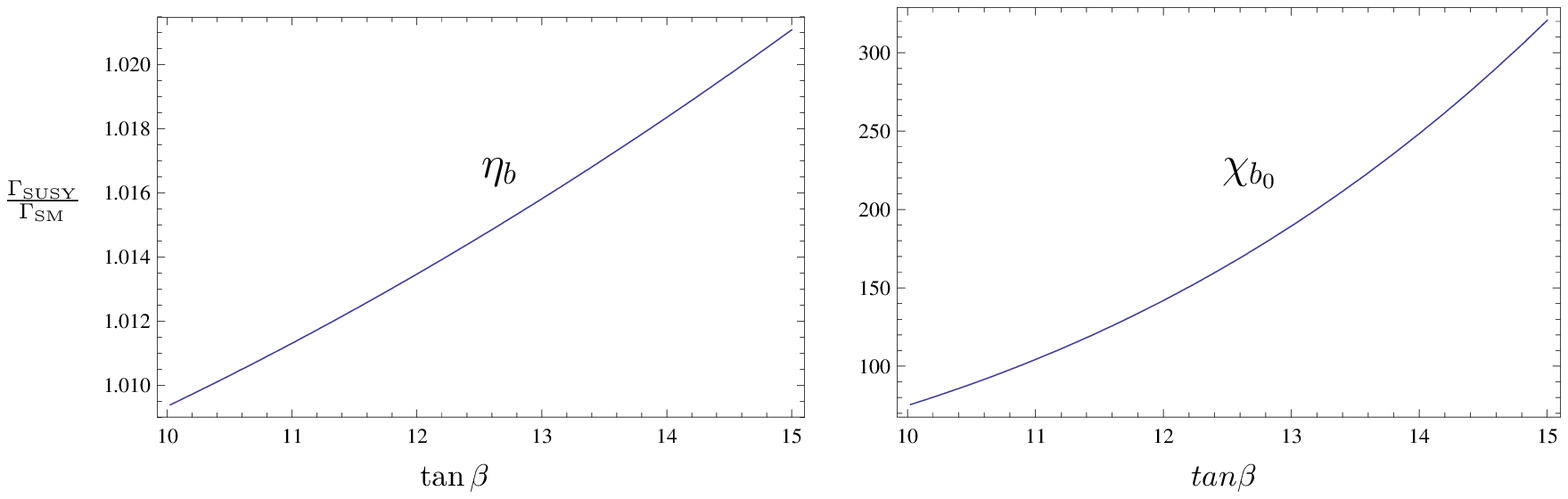}\hspace{1.5 cm}
    \vspace{-.5cm}
    \caption[] {The same as Fig.\ref{fig7}, but for $\texttt{SPSII}$ parameter set.
    }
    \label{fig8}
    \end{center}
\end{figure}

Another important parameter for the MSSM is the ratio of the vacuum
expectation values of the Higgs bosons, the $\tan\beta$. The
$\tan\beta$ dependencies of  the related decays are depicted in
Figs. \ref{fig7} and \ref{fig8} for $\eta_b$ and $\chi_{b 0}$ for
the two parameter sets employed in previous figures.

For the $\tan\beta$ values contained in $\texttt{SPSI}$ and
$\texttt{SPSII}$, $\eta_b$ decay exhibits less sensitivity to
$\tan\beta$ compared to $\chi_{b 0}$ decay. Nevertheless, the
$\eta_b$ decay stands as a sensitive probe of $\tan\beta$ since it
scales as $\left(\tan\beta\right)^4$, which becomes quite sizeable
at large values of $\tan\beta$. For instance, as can be seen from
the left panel of Fig. \ref{fig7} MSSM's prediction can be $\sim
1.5$ times larger than of the SM for reasonable values of
$\tan\beta$. The impact of the $\tan\beta$ variable for the $\chi_{b
0}$ decays is always supportive to claim that the MSSM prediction
can be four (right panel of Fig. \ref{fig7}) or two orders (right
panel of Fig. \ref{fig8}) of magnitude larger than the SM results.

\begin{figure}[htb]
\begin{center}
\vspace{0.5cm}
    \includegraphics[height=5.5cm,width=16cm,angle=0]{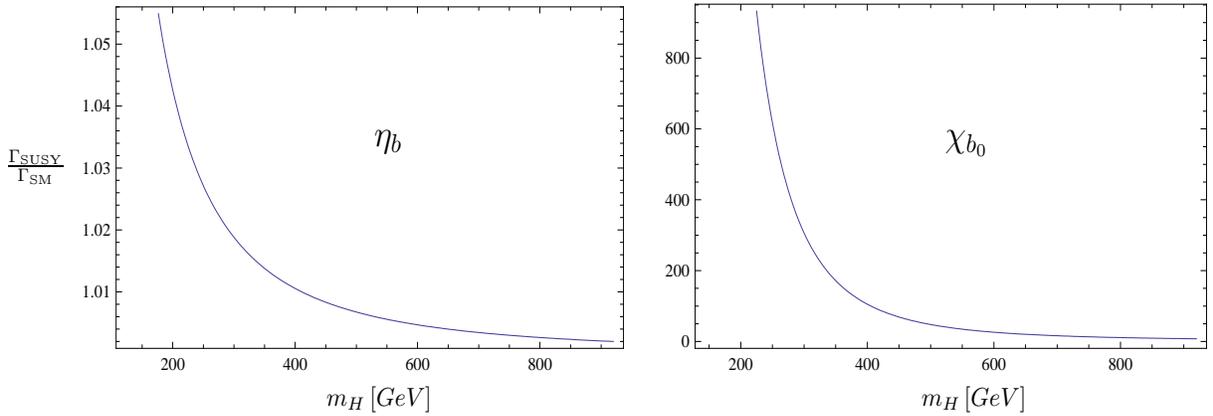}\hspace{1.5 cm}
    \vspace{-.5cm}
    \caption[]{Variations of the decay rate ratios against $m_H=m_A$. Here we
    fix the parameters as $\tan\beta=10$ and $m_h=120\rm{\,GeV}$.
    }
    \label{fig9}
    \end{center}
\end{figure}

Our last figure is devoted to examining the decay rates in the
decoupling limit {\it i.e.} the domain in which $m_A = m_H$ and it
is much larger than $m_h$. For this aim, we take $\tan\beta=10$,
$m_h=120\ {\rm GeV}$ and vary  $m_H = m_A$ from $124$  to $920\ {\rm
GeV}$. The numerical results are depicted in Fig. \ref{fig9}. As can
be seen from the left panel of the very figure, $\eta_b$ ratio
decreases as $m_H=m_A$ increases and its prediction does not offer a
difference more than $\sim 3 \%$. The largest effect occurs when
$m_H=m_A$ is not much larger than $m_h$, which actually means that
the $\eta_b$ decay cannot give any significant result in the
decoupling regime.

Coming to $\chi_{b 0}$, however, one notes from the right panel of
Fig. \ref{fig9} that, $\chi_{b0}$ is very sensitive to the variation
of $m_A = m_H$, especially for low values of the heavy Higgs bosons.
As $m_H$ converges to $m_h$ the ratio
$\Gamma^{\rm{MSSM}}\left(\chi_{b 0}\rightarrow \ell^+
\ell^-\right)/\Gamma^{\rm{SM}}\left(\chi_{b 0}\rightarrow \ell^+
\ell^-\right)$ can be enhanced up to $\sim 900$. Of course this
ratio can be further enhanced by increasing the $\tan\beta$. The
lesson from this figure is that the $\chi_{b 0} \rightarrow \ell^+
\ell^-$ decay in supersymmetry is a candidate with significantly
enhanced predictions with respect to the standard model rate.

\begin{table}
\begin{center}
\begin{tabular}{ccccc}
\hline\hline
&&\\[-0.2cm]
Decay ~~~~& ~~~~Potential~~ &~~~~~~~~   SM  ~~~~~~~~&~~~~~MSSM(SPSI)~~~~~& ~~~~MSSM(SPSII)\\
[0.2cm] \hline
&&\\[-0.2cm]
$\eta_b \rightarrow \ell^+ \ell^-$ & Cornell    & $5.08\times10^{-7}$ & $6.23 \times10^{-7}$& $5.13 \times10^{-7}$\\
                                   & Richardson & $2.37\times10^{-7}$ & $2.91 \times10^{-7}$& $2.39 \times10^{-7}$\\
                                   & Wisconsin & $1.86\times10^{-7}$ & $2.29 \times10^{-7}$& $1.88 \times10^{-7}$\\
                                   & Coulomb    & $1.02\times10^{-7}$ & $1.25 \times10^{-7}$& $1.03 \times10^{-7}$\\

\hline\hline
\end{tabular}
\end{center}
\caption{The branching ratios of the $\eta_b$ decay for different
potential models \cite{barger1} in the SM and in the MSSM.}
\label{tab1}
\end{table}

\begin{table}
\begin{center}
\begin{tabular}{ccccc}
\hline\hline
&&\\[-0.2cm]
Decay ~~~~& ~~~~Potential~~ &~~~~~~SM(SPSI, SPSII)~~~~~~&~~~~MSSM(SPSI, SPSII)~~~~\\
[0.2cm] \hline
&&\\[-0.2cm]
$\chi_{b0}\rightarrow \ell^+ \ell^-$ & Coulomb & ($6.20\times10^{-16}$, $3.25 \times10^{-16}$) ~~~&~~~ ($4.03 \times10^{-12}$, $2.36 \times10^{-14}$)\\
& others  & ($6.20\times10^{-14}$, $3.25 \times10^{-14}$) ~~~&~~~ ($4.03 \times10^{-10}$, $2.36 \times10^{-12}$)\\
\hline\hline
\end{tabular}
\end{center}
\caption{The branching ratios of the $\chi_{b0}$ decay as in Tab.
\ref{tab1}.} \label{tab2}
\end{table}

Moreover, we estimated the branching ratios of the $\eta_b$ and
$\chi_{b0}$ decays into $\ell^+\ell^-$ pairs for both SM and MSSM
processes using two parameter spaces: SPSI and SPSII. In doing this,
different potential model wave functions are examined \cite{barger1}
to probe the arbitrariness in the potential dependency. Our findings
are presented in Tabs. \ref{tab1} and \ref{tab2} for  $\eta_b$ and
$\chi_{b0}$ decays, respectively. As input parameters  we used
$\Gamma_{\eta_b}=10$ MeV taken from \cite{ref17} and
$\Gamma_{\chi_{b0}}=320$ keV  from \cite{ref18}.

As can be read from Tab. \ref{tab1} our predictions for the
branching ratios of the $\eta_b \rightarrow \ell^+ \ell^-$ decay in
the MSSM is $\sim$ 1 (SPSII) and 1.2 (SPSI) times larger than the SM
values. On the other hand, as can be read from Tab. \ref{tab2}, MSSM
predictions for the $\chi_{b0}\rightarrow \ell^+ \ell^-$ branching
ratio  can be as large as  ~73 (SPSII)  or even ~6500 (SPSI) times
larger than that of the SM predictions .

It should be noticed for both of the decays that they are rare
decays. It is possible to enhance the related predictions
theoretically in the MSSM, as examined in this section, but
experimental verification of such predictions is a challenging task.
\section{Conclusion}

In this work we have studied dileptonic bottomonium decays in
regard to their sensitivity to Higgs bosons of either CP
quantum number. We have found that, dileptonic branching of
$\chi_{b 0}$ is a highly sensitive probe of the extended Higgs
sector in that the rate increases significantly compared to the
SM prediction.

Theoretically, comparison of the $\eta_b$ ratio with the $\chi_{b
0}$ ratio shows that, for the selected parameter ranges, the
likelihood of observing the Higgs bosons via dileptonic $\eta_b$
decays turns out to be much smaller than those of $\chi_{b 0}$
decays. On the other, experimentally, since the predictions of the
branching ratios are at the order of  $\sim 10^{-7}$ for the $\eta$
and $\sim10^{-10}$ for the $\chi$ decays, both in the SM and in the
MSSM, $\eta_b$ turns out to be a better candidate for the
observation of the  Higgs bosons over these rare decays.

The results found here, given the high-luminosity, high-energy
nature of the LHC experiments, can be tested at the LHCb experiments
is not at the $B$ factories. Such a test, if conducted, would
provide a confirmation strategy if not a discovery strategy for
extended Higgs sectors. The recent paper by \cite{Domingo:2008rr}
also discusses the bottomonium decays with particular emphasis on
light pseudoscalar Higgs which can be realized in the NMSSM.

\section{Acknowledgments} I would like to thank to D. A.
DEM{\.{I}}R for his contributions with inspiring and illuminating
discussions in various stages of this work.

\end{document}